\documentclass[letterpaper,11pt]{article}
\usepackage{tabularx}
\usepackage{amsmath} 
\usepackage{graphicx}
\usepackage[margin=1in,letterpaper]{geometry}
\usepackage{hyperref}
\hypersetup{colorlinks,allcolors=black}
\usepackage{caption}
\usepackage{comment}
\usepackage{siunitx} 
\usepackage{chemformula}
\usepackage[style=numeric,url=false,maxbibnames=20,isbn=false,sorting=none]{biblatex}
\begin{document}

\title{\textbf{A Nanocryotron Memory and Logic Family}}
\author{Alessandro Buzzi$^1$, Matteo Castellani$^1$, Reed A. Foster$^1$, Owen Medeiros$^1$,\\ Marco Colangelo$^1$, Karl K. Berggren$^1$ \\[1ex]
\emph{$^1$Department of Electrical Engineering and Computer Science,} \\
\emph{Massachusetts Institute of Technology, Cambridge, MA, USA.}}
\date{\today}
\maketitle

\begin{abstract}
The development of superconducting electronics based on nanocryotrons has been limited so far to few-device circuits, in part due to the lack of standard and robust logic cells.
Here, we introduce and experimentally demonstrate designs for a set of nanocryotron-based building blocks that can be configured and combined to implement memory and logic functions.
The devices were fabricated by patterning a single superconducting layer of niobium nitride and measured in liquid helium on a wide range of operating points.
The tests show $10^{-4}$ bit error rates with above $\SI{20}{\percent}$ margins up to $\SI{50}{\mega\hertz}$ and the possibility of operating under the effect of a perpendicular $\SI{36}{\milli\tesla}$ magnetic field, with $\SI{30}{\percent}$ margins at $\SI{10}{\mega\hertz}$. 
Additionally, we designed and measured an equivalent delay flip-flop made of two memory cells to show the possibility of combining multiple building blocks to make larger circuits.
These blocks may constitute a solid foundation for the development of nanocryotron logic circuits and finite-state machines with potential applications in the integrated processing and control of superconducting nanowire single-photon detectors.
\end{abstract}

\vspace{5ex}

The nanocryotron (or nTron) is a superconducting three-terminal component that was proposed as a candidate for low-power electronics in 2014 \cite{mccaughan_superconducting-nanowire_2014}.
This device has gathered increasing attention thanks to its unique properties among superconducting devices, such as the capability of driving high-impedance loads and the possibility of operation in ambient magnetic fields \cite{mccaughan_superconducting-nanowire_2014}.
Several proof-of-concept circuits based on nanocryotrons were demonstrated, including a memory cell\cite{butters_scalable_2021}, a binary encoder \cite{zheng_superconducting_2020}, and an artificial neuron for neuromorphic computing \cite{toomey_superconducting_2020}.
In addition, the possibility of interfacing the nTron with other technologies, such as superconducting nanowire single-photon detectors (SNSPDs), rapid single flux quantum (RSFQ), and CryoCMOS, has been demonstrated \cite{zhao_nanocryotron_2017}.
Interfacing these technologies is beneficial towards the development of integrated sensors and superconducting optoelectronic hardware \cite{shainline_superconducting_2019}.

Despite the promising characteristics, nTron-based circuits have just been developed at scales of few devices, limiting their potential applications.
In particular, the design of larger circuits has been hindered by the absence of nTrons standard cells that can be combined together.
Therefore, the purpose of this work is to introduce a set of building blocks that integrate memory and logic functions and demonstrate them experimentally, in isolation as individual cells, and cascaded to show the feasibility of large-scale logic circuits.

\begin{figure}[t]
\includegraphics[width=0.9\textwidth]{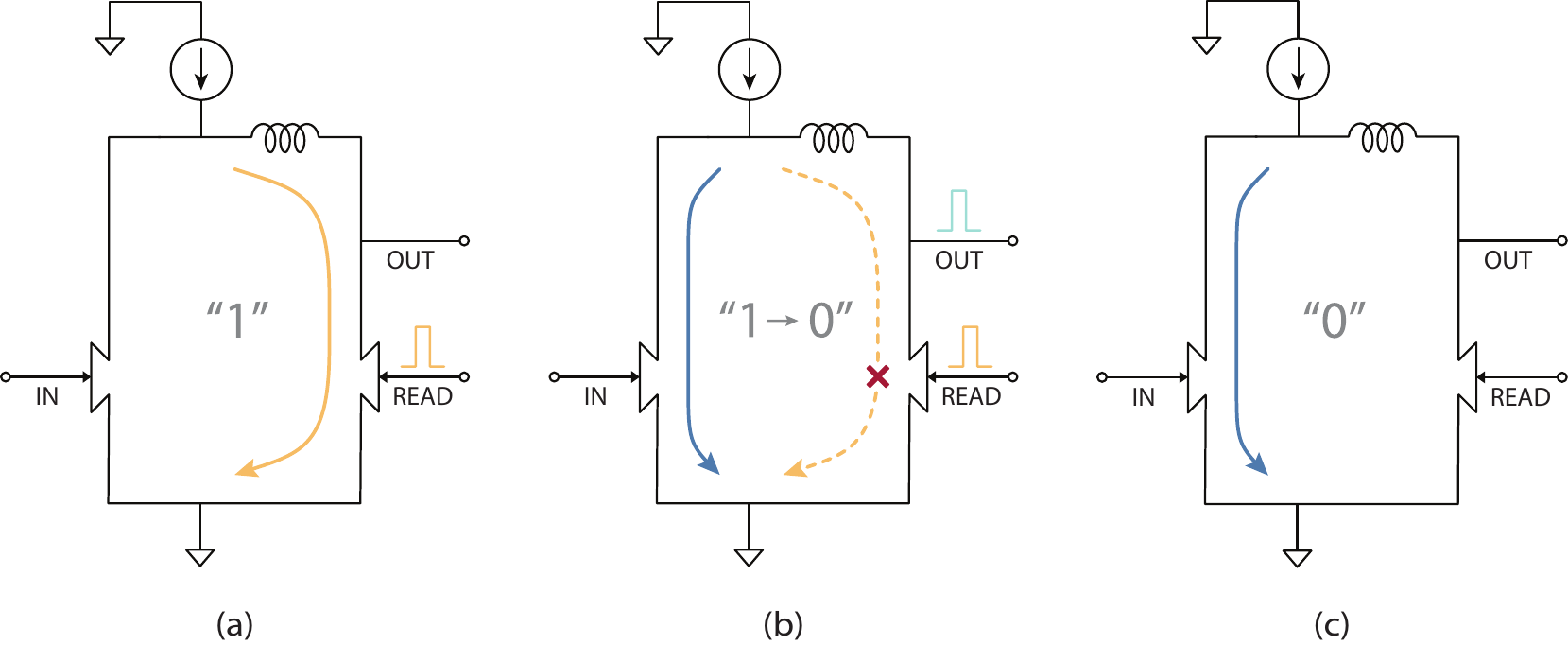}
    \centering
    \caption{Memory cell reading procedure.
    (a) The memory cell is in the “1” state, the current flows in the right branch.
    The cell can be read by sending a spike to the READ terminal.
    (b) When the spike arrives, a hotspot develops in the corresponding nTron channel.
    The resistive region produces a voltage spike at the OUT terminal and moves the bias current to the left branch.
    (c) The current returns to flow in the left branch.
    The cell is reset to the “0” state.}
    \label{fig:read}
\end{figure}

In the following, we describe the elementary cell of the logic family, which is a destructive readout memory.
The cell, shown in  Fig. \ref{fig:read}, is composed of a superconducting loop, a kinetic inductor, and two nTrons, whose gates are connected to the input (IN) and read (READ) terminals.
The output terminal (OUT) is placed above the READ.
A direct current biases the loop. The information is stored in the persistent current, flowing in one or the other branch of the loop, determining the state of the memory.
This state can be modified by sending current spikes to the nTrons’ gate terminals.
The gate of an nTron is connected to the device’s main channel by a narrow choke.
If the current exceeds the critical current of the choke, a localized resistive area, or hotspot, is generated at the gate constriction.
The presence of this hotspot suppresses the critical current of the corresponding branch of the loop.
Therefore, if the bias current is flowing in that branch, the whole nTron channel switches to the normal state, diverting the bias current to the opposite branch of the loop.

Due to the presence of the kinetic inductor in the right branch of the loop, when the cell is turned on, the bias current flows almost entirely into the left branch.
This state, depicted in Fig. 1c, corresponds to the zero state (“0”).
The state is left unaltered if a READ signal is sent because the right branch is not biased, so no output signal is generated.
However, if a current spike is injected into the input (IN) gate, a hotspot in the left branch of the loop is generated, and the bias current is transferred to the right branch.
The memory state switches to “1” (Fig. \ref{fig:read}a).
Now, if the cell is read (Fig. \ref{fig:read}b) by sending a current spike to the READ terminal, a voltage spike is produced at the output terminal (OUT), and the current is moved back to the left branch of the cell.
The state is reset to “0” (Fig. \ref{fig:read}c).
In other words, the presence or absence of an output voltage spike depends on the binary state of the loop, when a READ signal is applied.

The device resembles an RSFQ set-reset latch in topology and behavioral functioning \cite{likharev_rsfq_1991}.
Unlike Josephson junctions, the nTron is a three-terminal device with a gate terminal that controls the state of the main channel.
This geometric difference allows multiple nTrons in the same superconducting loop to be addressed individually.
By adding extra nTrons to the memory cell, different logic gates can be obtained using a single loop, which is not the case for RSFQ gates.

\begin{figure}[t]
\includegraphics[width=0.9\textwidth]{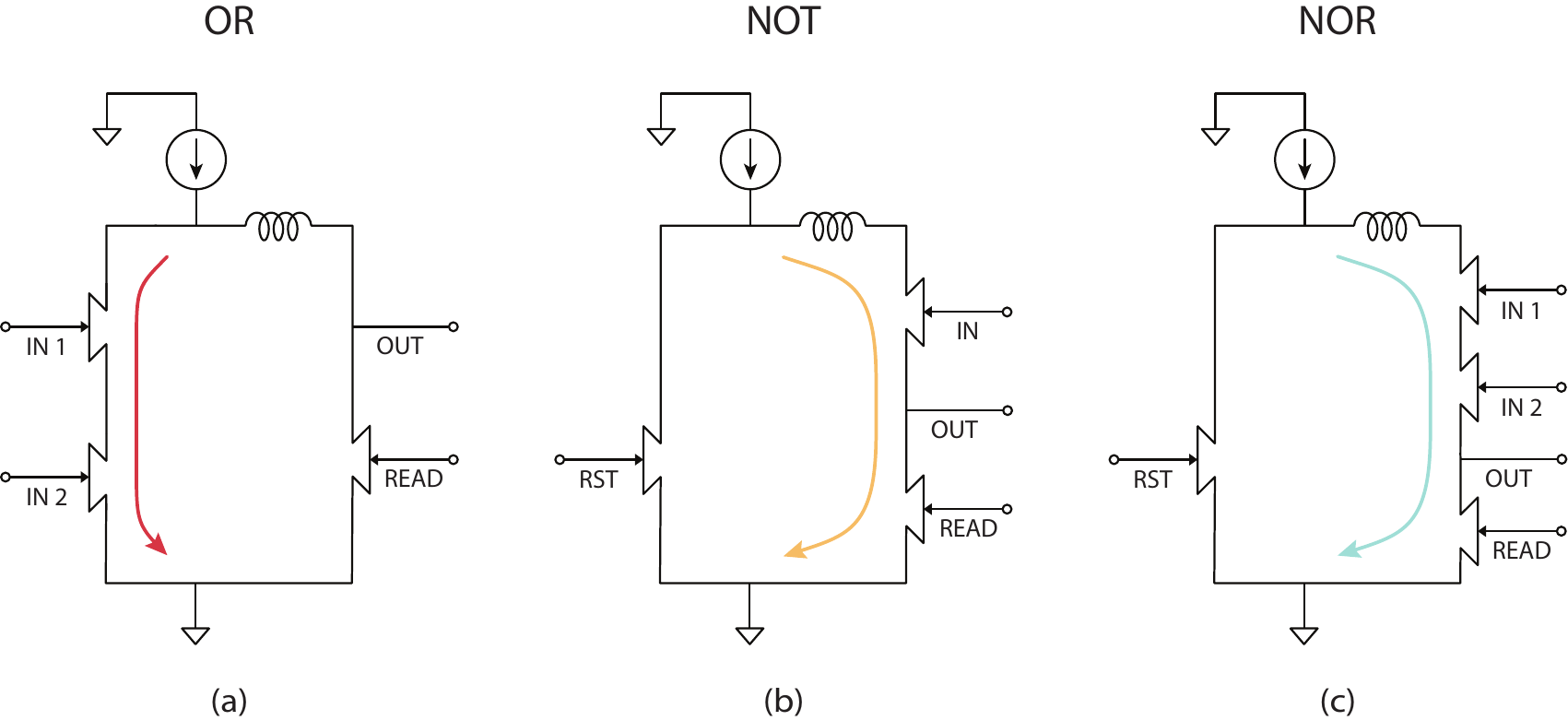}
    \centering
    \caption{Logic gates circuit schematic.
    (a) If extra nTrons are added to the left branch of the loop, the cell switches to the “1” state if any of them receives an input spike.
    Therefore, the device behaves as an OR gate.
    (b) A NOT gate can be obtained by moving the input to the right branch and adding a reset nTron in the left branch to precharge the state to “1”.
    (c) A NOR gate can be made by adding extra inputs to the NOT gate.
    After reset, any input spike would move the current to the left branch and, thus, the state to “0”.
    For the sake of simplicity, the shunt and output resistors (connected to the top of the left branch and the OUT terminal) are not shown in the schematics.}
    \label{fig:conf}
\end{figure}

Fig. \ref{fig:conf} illustrates three such gates.
Fig. \ref{fig:conf}a shows an OR gate, which is created by adding an extra input to the left branch of the memory cell.
If either input receives a current spike, the bias current is diverted to the right branch, setting the cell to the “1” state.
To make a NOT gate, the input terminal is moved to the right branch, and an extra reset (RST) nTron is inserted in the left branch (Fig. \ref{fig:conf}b).
In this case, the RST nTron is switched first, to preset the cell state to “1”.
If a current spike is injected into the IN terminal before reading, the cell is set to “0”.
Otherwise, the cell remains in the “1” state.
Lastly, if extra inputs are added to the NOT gate, the cell will be set to “0” if any of these receives a current spike.
The cell behaves as a NOR gate (Fig. \ref{fig:conf}c).
Since the NOR operation is functionally complete (i.e., every logic operation can be expressed in terms of NORs), the set of these gates is logically universal.
It is worth noticing that in the inverting gates, the output terminal is placed above the READ but below the inputs.
Otherwise, each input pulse would generate unwanted output signals due to the resistive state generation.

\begin{figure}[t]
\includegraphics[width=0.9\textwidth]{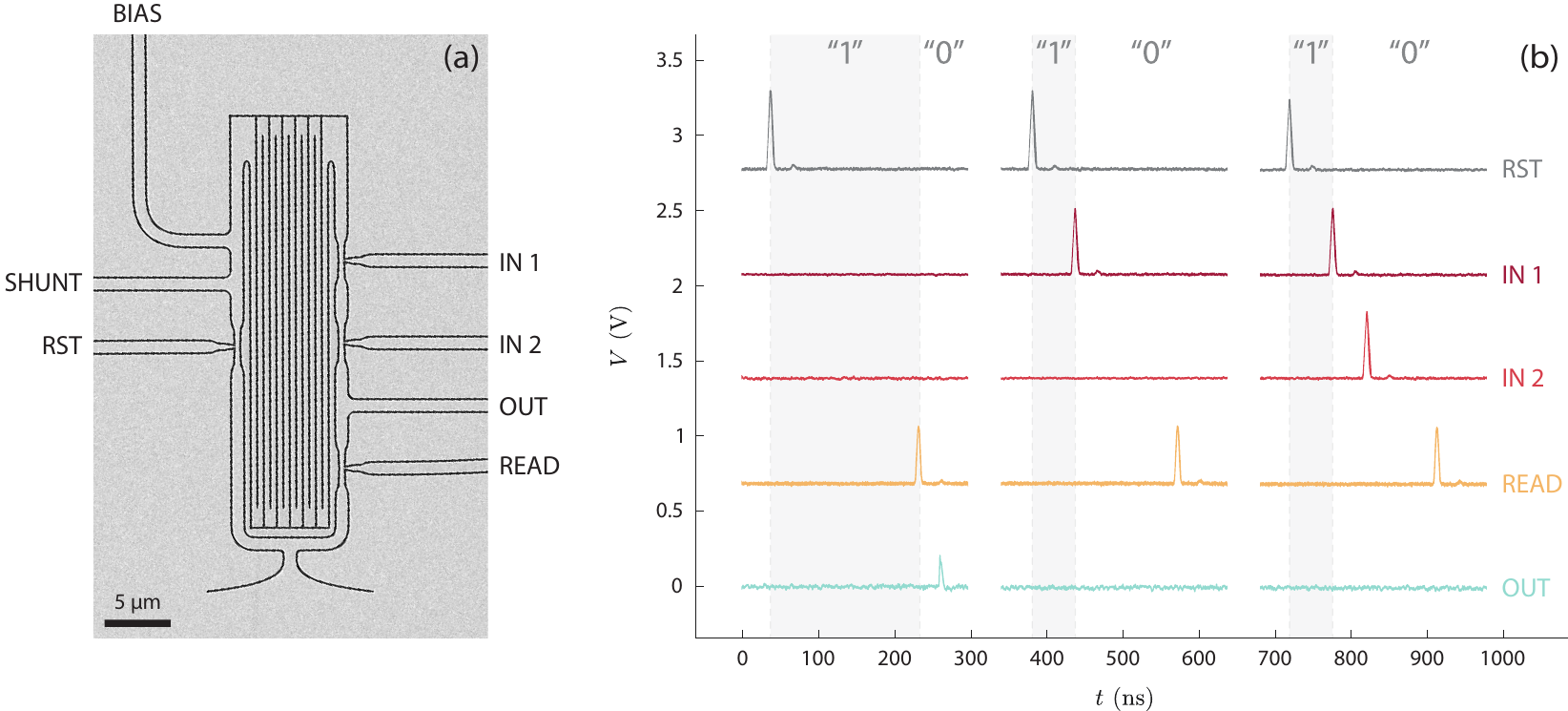}
    \centering
    \caption{NOR gate micrograph and experimental voltage traces.
    (a) Scanning electron micrograph of a NOR gate with a 20 nH kinetic inductor, in the middle of the loop, and all the electrical terminals.
    (b) Experimental voltage traces of a NOR gate with different input configurations.
    In the three cases, the cell is preset to “1” by sending a current spike to the RST terminal.
    When no input arrives, in the first case, the state of the cell is preserved until reading.
    The read spike generates a corresponding output spike (the delay between the reading spike and the output signal is mainly given by the wiring of the setup and not by internal effects).
    Conversely, when one or both inputs fire, in the second and third cases, the cell is set back to “0”, and no output is generated.}
    \label{fig:nor}
\end{figure}

We determined the circuit parameters of the devices with SPICE simulations, using an electrical model of the nTron \cite{castellani_design_2020, berggren_superconducting_2018}.
Starting from the obtained values, the devices were fabricated on silicon dioxide on silicon chips, with an oxide thickness of $\SI{300}{\nano\meter}$.
A $\SI{15}{\nano\meter}$-thick niobium nitride (\ch{NbN}) film was deposited by sputtering \cite{dane_bias_2017}, patterned with electron beam lithography, and etched with \ch{CF4} reactive ion etching.
A scanning electron micrograph of a NOR gate with a $\SI{20}{\nano\henry}$ inductor is depicted in Fig. \ref{fig:nor}a.
Finally, the chip was wire-bonded to a printed circuit board, on which $\SI{50}{ohm}$ shunt resistors were soldered.
The circuits were tested in liquid helium ($\SI{4.2}{\kelvin}$).
In this work, we experimentally tested multiple devices with different values of loop inductance: five memory cells with $\SI{60}{\nano\henry}$, $\SI{40}{\nano\henry}$, $\SI{20}{\nano\henry}$, and $\SI{5}{\nano\henry}$,  two OR gates with $\SI{60}{\nano\henry}$ and $\SI{40}{\nano\henry}$, two NOR gates with $\SI{20}{\nano\henry}$, and a NOT gate with $\SI{20}{\nano\henry}$.
All the devices’ nTrons had a $\SI{300}{\nano\meter}$-wide main channel and a $\SI{30}{\nano\meter}$-wide choke.

The experimental voltage traces for different input configurations of a NOR gate are presented in Fig. \ref{fig:nor}b.
In the first section of the plot, after the cell is set to “1” by the RST signal, no input arrives.
Therefore, when the reading spike occurs, an output pulse is generated (the state “1” is read out).
For the second and third sections of the plot, a current spike is injected into an input terminal, moving the bias current to the left branch, and thus the memory is set to the “0” state.
Therefore, if one of the inputs or both of them fire before the READ spike, no output pulse is generated, as expected for a NOR gate.

\begin{figure}[t]
\includegraphics[width=0.9\textwidth]{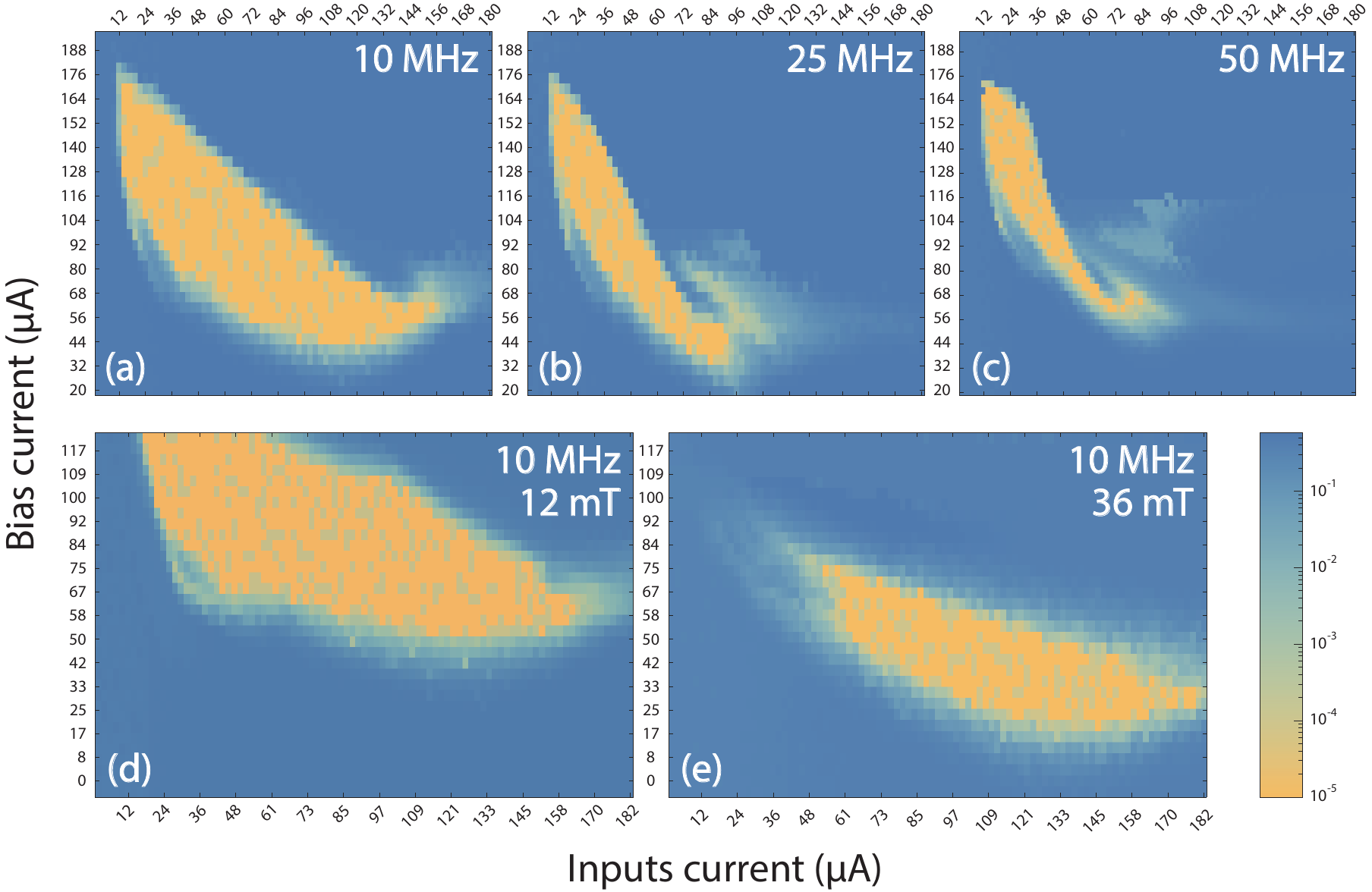}
    \centering
    \caption{Memory cell bit error rate (BER) at different frequencies and magnetic field intensities.
    The BER is evaluated on a range of operating points at (a) $\SI{10}{\mega\hertz}$, (b) $\SI{25}{\mega\hertz}$, and (c) $\SI{50}{\mega\hertz}$, and at $\SI{10}{\mega\hertz}$ under a magnetic field of (d) $\SI{12}{\milli\tesla}$, and (e) $\SI{36}{\milli\tesla}$.
    Each bias point was measured on $10^4$ pseudo-randomly generated input bits.}
    \label{fig:ber}
\end{figure}

We measured the bit error rate (BER) of a memory cell (with a $\SI{40}{\nano\henry}$ inductor) to characterize the performance of the devices.
The analysis took into account a range of bias and input currents to study the margins of operation of the circuit.
Each point of the sweep was tested on a $10^4$ pseudo-random stream of bits.
Furthermore, The BER, shown in Fig. \ref{fig:ber}, was evaluated at different frequencies and in a magnetic field.
Figs. \ref{fig:ber}a-c show the variation of the low-BER area (in yellow) as a function of frequency.
At $\SI{10}{\mega\hertz}$, the operating point margins, from the center of the region, are above $\SI{40}{\percent}$ in both directions.
With the increase in frequency, the area gets narrower in the horizontal direction, corresponding to the peak current of the input spikes.
The variation is related to a more frequent occurrence of latching.
The device does not work either if the currents are too low to generate a hotspot (under-biasing) or high enough to make the device latch.
The second phenomenon arises when the main channel of the device does not recover to the superconducting state \cite{kerman_electrothermal_2009}.
As a result, the nanocrytron latches to the resistive state until the bias current is removed.
If nTrons in different branches switch simultaneously, both of them may latch, preventing proper functioning.
To reduce this effect, input and read signals must be separated by a setup time, limiting the maximum frequency of operation.
The device was successfully tested up to $\SI{100}{\mega\hertz}$, but with very narrow margins on the operating point, due to latching.

One of the advantages of nTrons over other superconducting technologies is the ability to work under the effect of a magnetic field \cite{mccaughan_superconducting-nanowire_2014}.
Therefore, the BER was measured at $\SI{10}{\mega\hertz}$ in the presence of a static magnetic field.
The corresponding results are shown in Figs. \ref{fig:ber}d-e.
With a $\SI{12}{\milli\tesla}$ field applied perpendicular to the chip, the BER remains essentially unaffected.
In contrast, if the field is increased up to $\SI{36}{\milli\tesla}$, the correct-functioning area shrinks and shifts to lower values of bias current.
This result is likely to be related to the extra current added to the loop by flux trapping \cite{polyakov_flux_2007}, which causes a decrease in the effective switching current of the branches.
Despite the shift, the device continues to perform with $\SI{30}{\percent}$ margins on the operating point.

The choice of the operating point also affects the energy per operation.
An upper bound to the energetic consumption can be estimated considering the bias current to flow for $\SI{5}{\nano\second}$ through a $\SI{50}{\ohm}$ resistance (strictly higher than the parallel combination of the hotspot resistance and the shunt) per switching event.
Depending on the current values, the energy per operation is approximately~$\SI{1}{\femto\joule}$.

\begin{figure}[t!]
\includegraphics[width=0.6\textwidth]{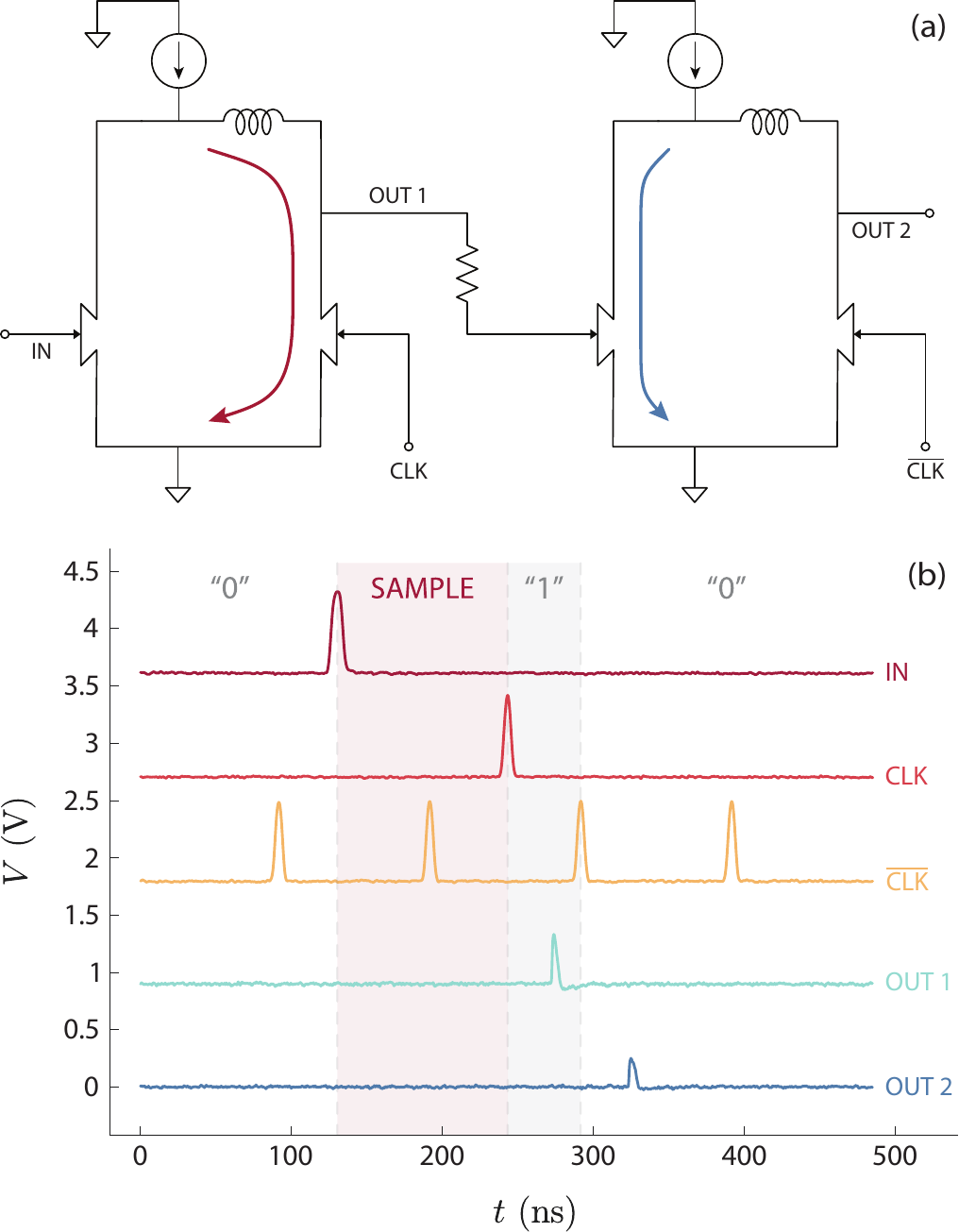}
    \centering
    \caption{Equivalent delay flip-flop schematic and experimental voltage traces.
    (a) The schematic shows the resistive connection of two memory cells controlled by out-of-phase clock signals.
    (b) Experimental voltage traces of two memory cells (with $\SI{60}{\nano\henry}$ inductors) connected together.
    When a current spike is sent to the IN terminal, the input is sampled, and the first cell is set to “1”.
    The value is transmitted to the second cell by a CLK spike while the first cell is reset.
    Finally, a $\overline{\textrm{CLK}}$ reads the second cell, resetting the device and producing an output (OUT 2 terminal) signal.}
    \label{fig:dff}
\end{figure}

After demonstrating and analyzing single-cell devices, we designed an equivalent master-slave delay flip-flop, made of two memories, to show the possibility of combining multiple cells.
The schematic of the circuit is reported in Fig. \ref{fig:dff}a.
In this flip-flop, each cell stores its logic state until reading. When a cell is read, the state is transferred to the following memory.
The reading of the two cells is performed by the out-of-phase signals CLK and $\overline{\textrm{CLK}}$.
The output of the first cell and the input of the second one are connected by a $\SI{20}{\ohm}$ resistor.
The presence of the resistance prevents any persistent current from being stored in the connection, which could cause unwanted switchings of the second cell input nTron.
The corresponding experimental voltage traces are displayed in Fig. \ref{fig:dff}b.
Initially, both memories are in the “0” state.
Therefore, any reading operation will not produce an output signal.
Once a current spike is sent to the IN terminal, the first cell switches to “1”.
When this cell is read with a CLK signal, the memory is reset, and a voltage spike is produced at the OUT 1 terminal.
This spike sets the second cell to the “1” state.
Its state can be read and reset by sending a current spike to the $\overline{\textrm{CLK}}$ terminal.
The circuit demonstrates the possibility of connecting multiple cells.
Moreover, it constitutes a basic element for other circuits, such as a shift register.

Following the analysis of the experimental results, it is worth examining further advantages and limitations involved in the scaling of this technology.
The main limitation for shrinking the size of the device is given by the kinetic inductor.
This component plays three fundamental roles in the cells’ functioning: (1) it sets the cell state to “0” when the device is turned on, (2) it increases the output impedance, and (3) it reduces the impact of latching.
All three points are favored by larger inductances.
Of particular concern is the output impedance, which affects the peak voltage of the output spikes.
Indeed, the smallest memory tested ($\SI{5}{\nano\henry}$ inductor) worked properly but produced lower signal levels.
Nevertheless, the inductor size problem could be solved by employing high fan-in gates in logic circuits.
The cost, in terms of area, of additional nTrons in the loop is negligible compared to the kinetic inductor area, which does not increase for multi-input gates.
The possibility of making multiple input gates, potentially with a larger fanout thanks to the high nTron’s output current \cite{mccaughan_superconducting-nanowire_2014}, could reduce the number of cells per logic circuit, and, thus, the overall area footprint.

In this work, we introduced a logic family based on nTrons which integrates combinatorial and sequential functions.
We demonstrated the operation of the devices experimentally, characterized the bit error rate under different conditions, and proved the possibility of connecting multiple cells together.
To further assess the technology's applicability, future work should focus on the demonstration of multiple gates logic circuits and integration with SNSPDs for elementary control and processing.

\section*{Acknowledgments}
The initial stages of this work were supported by the DARPA Invisible Headlights program.
The final stage of the work (including data analysis and manuscript preparation) was supported by the DOE.
We would like to thank James Daley and Mark Mondol of the NanoStructures Laboratory for technical assistance.

\section*{Data Availability}
The data that support the findings of this study are available from the corresponding author upon reasonable request.

\printbibliography


\end{document}